\documentclass[twocolumn]{article}
\usepackage{amsmath,amssymb,amsthm}
\usepackage{algorithm,algorithmic,multirow}
\usepackage{subcaption,graphicx,url}
\title{Unsupervised Adversarial Domain Adaptation Based On The\\Wasserstein Distance For Acoustic Scene Classification}
\author{Konstantinos Drossos, Paul Magron, and Tuomas Virtanen\\
Audio Research Group, Tampere University, Tampere, Finland}
\date{}
\begin{document}
%
\twocolumn[
\maketitle
  \begin{@twocolumnfalse}
    \maketitle
    \begin{abstract}
      A challenging problem in deep learning-based machine listening field is the degradation of the performance when using data from unseen conditions. In this paper we focus on the acoustic scene classification (ASC) task and propose an adversarial deep learning method to allow adapting an acoustic scene classification system to deal with a new acoustic channel resulting from data captured with a different recording device. We build upon the theoretical model of $\mathcal{H}\Delta\mathcal{H}$-distance and previous adversarial discriminative deep learning method for ASC unsupervised domain adaptation, and we present an adversarial training based method using the Wasserstein distance. We improve the state-of-the-art mean accuracy on the data from the unseen conditions from 32\% to 45\%, using the TUT Acoustic Scenes dataset.

\textbf{Keywords:} Acoustic scene classification, unsupervised domain adaptation, Wasserstein distance, adversarial training
    \end{abstract}
    \vspace{18pt}
  \end{@twocolumnfalse}
]
%
%
\section{Introduction}
The task of acoustic scene classification (ASC) consists in classifying a sound segment $\mathbf{x}\sim\mathcal{X}$ into one of the predefined classes $y\sim\mathcal{Y}$ representing different acoustic scenes (e.g., ``urban area'', ``metro station''), where $\mathcal{X}$ and $\mathcal{Y}$ are the underlying distributions of the sound segments and predefined classes, respectively. ASC has recently been tackled with deep learning methods~\cite{han:2017:dcase,weiping:2017:dcase,sakashita:2018:dcase,dorfer:2018:dcase}. These methods can be viewed as consisting of a feature extractor that outputs a latent representation $\mathbf{z}$ from $\mathbf{x}$ and a label classifier that assigns a label $\mathbf{y}$ to $\mathbf{z}$. 

Training and testing data can exhibit different capturing conditions (e.g., different recording devices) and subjective and noisy labels (e.g., an ``urban area'' scene labeled as ``residential area'' and ``town''), resulting in the degradation of the performance of the method~\cite{gharib2018unsupervised,cortes:2008:alt}. One promising way for tackling the above mentioned problems is \emph{domain adaptation} (DA)~\cite{ganin2016domain, torralba:2011:cvpr, gretton:2009:chapter, tzeng:cvpr:2017} and, in the relevant terminology, the above mentioned problems are the capture and label biases, respectively~\cite{torralba:2011:cvpr,Tommasi_GCPR_2015}. 

Domain adaptation is a sub-space alignment/divergence minimization problem which consists in optimizing a system on the data from the \emph{source domain} and then adapting it to the data from the \emph{target domain}~\cite{ganin2016domain, tzeng:cvpr:2017, ben-david:2010:mlrj}. The data from the source and target domains exhibit any combination of the capturing and labeling biases, leading to what is known as domain shift or dataset bias phenomenon. This phenomenon manifests as differences between the distributions $\mathcal{X}_{S}$ and $\mathcal{X}_{T}$, and/or $\mathcal{Y}_{S}$ and $\mathcal{Y}_{T}$, where the indices $S$ and $T$ refer to the source and target domains, respectively. The goal of DA is to minimize the error of a classifier when classifying data from the target domain. This is usually achieved by training a system which minimizes the discrepancy between the distributions $\mathcal{Z}_S$ and $\mathcal{Z}_T$ of the learned latent representations from the source and target domains, $\mathbf{z}_{S}$ and $\mathbf{z}_{T}$ respectively. All the labels for the source data are available but only some or none labels are available for the target data. This results in semi- or unsupervised DA, respectively. 

We consider here the unsupervised adversarial DA, focused on the capturing bias problem. Previous approaches can be organized in two categories, depending on when DA is performed. Methods in the first category such as~\cite{ganin2016domain,zhao2018multiple,pei:2018:aaai} perform DA jointly with the optimization on the source domain data. Methods in the other category such as~\cite{tzeng:cvpr:2017,hoffman:2018:jmlr,chadha:2018:arxiv} are two-stage approaches that first perform optimization on the source domain data and then do DA. The previous ASC unsupervised adversarial DA state-of-the-art (SOTA) method~\cite{gharib2018unsupervised} falls in this second category. However, this method suffers from vanishing gradients and slow learning process, during the minimization of the discrepancy between the distributions of the source and target domains. 

In this paper we present a deep learning method for ASC unsupervised adversarial DA which is based on the theoretical analysis of domain adaptation based on optimal transport, presented in~\cite{redko:2017:mlkdd}. We replace the adversarial adaptation process SOTA method~\cite{gharib2018unsupervised} with the Wasserstein generative adversarial networks (WGAN) formulation~\cite{arjovsky:2017:icml}, tackling the computational problems that hampered the performance of the previous SOTA method. Furthermore, we enhance the WGAN algorithm by using an extra loss for preventing the performance from dropping on the source domain when the system is adapted to the target domain. We achieve an increase of 13\% in the mean accuracy on the target data compared to previous SOTA approach~\cite{gharib2018unsupervised} using the TUT Acoustic Scenes dataset~\cite{gharib:2018zenodo-audasc}. The contributions of our work are: 
\begin{enumerate}
    \item Providing the first theoretical based approach for deep learning DA on general audio;
    \item Presenting the first work adopting the Wasserstein distance for ASC unsupervised adversarial DA;
    \item Significantly improving the performance of ASC unsupervised adversarial DA over the previous SOTA method~\cite{gharib2018unsupervised}.
\end{enumerate}
The rest of the paper is structured as follows. In Section~\ref{sec:background} we present the most widely used theoretical background in DA. In Section~\ref{sec:method} we introduce our proposed method. The setup for evaluating our method is presented in Section~\ref{sec:evaluation} and the results are discussed in Section~\ref{sec:results}. Finally, the paper is concluded in Section~\ref{sec:conclusions}.
%
%
\section{Background}\label{sec:background}
We present here a summary of the most widely used theoretical background for understanding adversarial deep learning based unsupervised DA methods, and the interested reader can refer to~\cite{ben-david:2010:mlrj,blitzer:2008:nips} for further details. 
%
%
\subsection{Classifier error and target error bound}\label{subsec:domainerror}
A domain $\mathcal{D}$ is defined as the pair of a distribution $\mathcal{Z}$ and a labeling process $f$. $\mathcal{Z}$ is the distribution of $\mathbf{z}$, i.e. $\mathbf{z}\sim\mathcal{Z}$, and $\mathbf{z}$ is used as an input to $f$. The output of $f$ is considered as the ground truth that a classifier $h$ should predict when performing classification over $\mathbf{z}$. It follows that $\mathcal{D}_{S}=\left<\mathcal{Z}_{S},f_{S}\right>$ and $\mathcal{D}_{T}=\left<\mathcal{Z}_{T},f_{T}\right>$ are the source and target domains, respectively. The expected error of an $h$ over its input $\mathbf{z}$ is
\begin{equation}\label{eq:error}
    \epsilon(h, f) = \mathbb{E}_{\mathbf{z}}[\mathcal{L}(h(\mathbf{z}),\,f(\mathbf{z}))]\text{,}
\end{equation}
\noindent
where $\mathcal{L}$ is a loss function. $\epsilon(h, f)$ indicates the average disagreement between the prediction of the classifier $h$ and the output of the labeling function $f$.

If $\mathbf{z}$ comes from the source (resp. target) domain, that is, $\mathbf{z}_{S}\sim\mathcal{Z}_{S}$ (resp. $\mathbf{z}_{T}\sim\mathcal{Z}_{T}$), then the error is the source error $\epsilon_{S}(h, f_{S})$ (resp. the target error $\epsilon_{T}(h, f_{T})$). The aim of unsupervised DA is to obtain a classifier $h$ that yields a low value for $\epsilon_{S}$ and adapt it to yield a low value for $\epsilon_{T}$, without employing the labels from $\mathcal{D}_{T}$ during the adaptation process. 

We can obtain a classifier $h$ that yields a low value for $\epsilon_{S}$ by following classical supervised training approaches. However, we cannot optimize $h$ on $\mathbf{z}_{T}$ from the target domain $\mathcal{D}_{T}$ in a supervised manner, since we do not have access to the labels from $\mathcal{D}_{T}$. Intuitively, if there was a low discrepancy between $\mathcal{Z}_{S}$ and $\mathcal{Z}_{T}$, then the classifier $h$ would also yield a low value for $\epsilon_{T}$. Calculating or estimating this discrepancy allows us to obtain a generalization bound for $\epsilon_{T}$, and reducing this bound
will consequently reduce $\epsilon_{T}$ as shown in~\cite{ben-david:2010:mlrj,redko:2017:mlkdd,vapnik:1999:springer}. 

The discrepancy between $\mathcal{Z}_{S}$ and $\mathcal{Z}_{T}$ can be measured with the $\mathcal{H}\Delta\mathcal{H}$-distance~\cite{ben-david:2010:mlrj,blitzer:2008:nips}. In the case of binary classifiers $\mathcal{H}$, the $\mathcal{H}\Delta\mathcal{H}$-distance is 
defined as
\begin{align}
    &d_{\mathcal{H}\Delta\mathcal{H}}(\mathcal{Z}_{S}, \mathcal{Z}_{T}) =\nonumber\\
    &2\underset{h,h'\in\mathcal{H}}{\sup}|\text{P}_{\mathbf{z}\sim\mathcal{Z}_{S}}[h(\mathbf{z})\neq h'(\mathbf{z})]-\text{P}_{\mathbf{z}\sim\mathcal{Z}_{T}}[h(\mathbf{z})\neq h'(\mathbf{z})]|\text{.}
  \label{eq:dhd}
\end{align}
\noindent
In a nutshell, $d_{\mathcal{H}\Delta\mathcal{H}}(\mathcal{Z}_{S}, \mathcal{Z}_{T})$ returns the highest prediction difference between two classifiers in $\mathcal{H}$ under the two distributions $\mathcal{Z}_{S}$ and $\mathcal{Z}_{T}$. We use $d_{\mathcal{H}\Delta\mathcal{H}}$ to bound the target error for DA~\cite{ben-david:2010:mlrj,blitzer:2008:nips,zhao2018multiple} as
\begin{equation}\label{eq:upper_bound}
    \epsilon_{T}(h, f_{T})\leq
    \epsilon_{S}(h, f_{S})+\frac{1}{2}d_{\mathcal{H}\Delta\mathcal{H}}(\mathcal{Z}_{S}, \mathcal{Z}_{T})+\lambda\text{,}
\end{equation}
where
\begin{align}
    \lambda = &\epsilon_{S}(h^{\star},f_{S}) + \epsilon_{T}(h^{\star},f_{T})\text{, and }\\
    h^{\star} = &\underset{h\in\mathcal{H}}{\text{argmin}}(\epsilon_{S}(h, f_{S}) + \epsilon_{T}(h, f_{T}))\text{.}
\end{align}
$h^{\star}$ and $\lambda$ are termed as the ideal joint classifier and combined error of the ideal joint classifier, respectively~\cite{ben-david:2010:mlrj,blitzer:2008:nips}. Eq.~\ref{eq:upper_bound} shows that the upper bound of the target error $\epsilon_{T}(h, f_{T})$ is affected by three factors: the error on the source domain (i.e., $\epsilon_{S}(h, f_{S})$); the $\mathcal{H}\Delta\mathcal{H}$-distance between the source and target distributions; and the combined error of the ideal joint classifier (i.e., $\lambda$). In DA it is safely assumed that there is a classifier that performs well on both the source and the target domains, thus yielding a small value of $\lambda$ and therefore allowing to neglect its effect~\cite{ben-david:2010:mlrj}. Under this assumption, the problem of DA reduces to obtaining a classifier with a good performance on the source domain (i.e., minimizing $\epsilon_{S}(h, f_{S})$) and trying to minimize the discrepancy between the distributions of the two domains (i.e., $d_{\mathcal{H}\Delta\mathcal{H}}(\mathcal{Z}_{S}, \mathcal{Z}_{T})$).
%
%
\subsection{Adversarial formulation}\label{subsec:adversarialform}
Consider a set of domain classifiers $\mathcal{H}_{d}$ (i.e., classifiers that predict whether $\mathbf{z}$ is from $\mathcal{D}_{S}$ or $\mathcal{D}_{T}$). Then, according to~\cite{ganin:2015:icml},
\begin{align*}
    &d_{\mathcal{H}\Delta\mathcal{H}}(\mathcal{Z}_{S}, \mathcal{Z}_{T}) =\nonumber\\
    &=2\underset{h,h'\in\mathcal{H}}{\sup}|\text{P}_{\mathbf{z}\sim\mathcal{Z}_{S}}[h(\mathbf{z})\neq h'(\mathbf{z})]-\text{P}_{\mathbf{z}\sim\mathcal{Z}_{T}}[h(\mathbf{z})\neq h'(\mathbf{z})]|\nonumber
\end{align*}
\begin{align}
    &\leq2\underset{h_{d}\in\mathcal{H}_{d}}{\sup}|\text{P}_{\mathbf{z}\sim\mathcal{Z}_{S}}[h_{d}(\mathbf{z})=1]-\text{P}_{\mathbf{z}\sim\mathcal{Z}_{T}}[h_{d}(\mathbf{z})=1]|\nonumber\\
    &=2\underset{h_{d}\in\mathcal{H}_{d}}{\sup}|\text{P}_{\mathbf{z}\sim\mathcal{Z}_{S}}[h_{d}(\mathbf{z}) = 0]+\text{P}_{\mathbf{z}\sim\mathcal{Z}_{T}}[h_{d}(\mathbf{z}) = 1] - 1|\text{.}\label{eq:dhd_bound}
\end{align}
\noindent
Thus, the ideal $h_{d}$ yields as an upper bound for $d_{\mathcal{H}\Delta\mathcal{H}}(\mathcal{Z}_{S}, \mathcal{Z}_{T})$~\cite{ganin2016domain}. This can be exploited in an adversarial setting, where the focus is to obtain a domain classifier $h_{d}$ good enough to predict the domain of $\mathbf{z}$ and a feature extractor that confuses the domain classifier. In this setting, instead of directly minimizing $d_{\mathcal{H}\Delta\mathcal{H}}$, one aims at obtaining a feature extractor $M$ that is able to fool a good domain classifier $h_{d}$.
%
%
\subsection{Previous SOTA}\label{subsec:previous-work}\vspace{-4pt}
The previous SOTA approach for ASC unsupervised adversarial DA~\cite{gharib2018unsupervised} falls in the second category of approaches mentioned in the introduction (i.e. first performing optimization on source domain and then doing DA). In the first stage, a feature extractor $M_{S}$ is obtained during the optimization of the label classification, and in the second stage, a copy $M_{T}$ of $M_{S}$ is further optimized during the adversarial training. Following the framework presented in Section~\ref{subsec:adversarialform}, the second stage of this approach consists of an adversarial training that aims at minimizing $d_{\mathcal{H}\Delta\mathcal{H}}(M_{S}(\mathbf{x}_{S}), M_{T}(\mathbf{x}_{T}))$. The implementation of the adversarial training follows the typical formulation initially presented in~\cite{goodfellow:2014:nips}, using two losses. The first one is used for optimizing $h_{d}$ over the output of $M_{T}$ and the second loss is utilized for making $M_{T}$ to produce an output that maximizes the domain classification error. In the field of generative adversarial neural networks (GANs), these are the losses associated with the discriminator (i.e., $h_{d}$) and the generator (i.e., $M_{T}$). The GAN loss $\mathcal{L}_{\text{GAN}}$ used in the previous SOTA approach can be formulated~\cite{gharib2018unsupervised,tzeng:cvpr:2017,goodfellow:2014:nips} as
\begin{align}
    \mathcal{L}_{\text{GAN}}(h_{d}, M) = \mathbb{E}_{\mathbf{x}_{S}}[\log \sigma(h_{d}(M_{S}(\mathbf{x}_{S})))] +\nonumber\\ \mathbb{E}_{\mathbf{x}_{T}}[\log(1-\sigma(h_{d}(M_{T}(\mathbf{x}_{T}))))]\text{,}\label{eq:gan-loss}
\end{align}
\noindent
where $\sigma$ is the sigmoid function. The adversarial process aims~\cite{goodfellow:2014:nips} at solving the optimization problem
\begin{equation}\label{eq:gan-minimax}
    \underset{h_{d}}{\min}\,\underset{M_{T}}{\max}\,\mathcal{L}_{\text{GAN}}(h_{d}, M_{T})\text{.}
\end{equation}
However, solving the optimization problem described in Eq.~\eqref{eq:gan-minimax} is known to introduce computational issues, such as vanishing gradients and slow learning process for $h_{d}$~\cite{arjovsky:2017:icml}. These problems hamper the performance of the adversarial training~\cite{arjovsky:2017:icml} and thus limit the performance of the method proposed in~\cite{gharib2018unsupervised}. The method proposed in this paper aims at tackling these problems for the ASC unsupervised DA task.
%
%
\section{Proposed method}\label{sec:method} 
Our proposed method builds upon the framework presented in Section~\ref{sec:background} and it can be applied with any deep neural network (DNN) that performs ASC. We employ a DNN and the Wasserstein generative adversarial networks (WGAN) formulation and algorithm, presented in~\cite{arjovsky:2017:icml}. Our DNN consists of a feature extractor $M$, a label classifier $h$, and a domain classifier $h_{d}$. We consider as $\mathbf{z}$ the output of our feature extractor $M$, which is used as an input to our $h$. The output of $h$ is a vector and the output of $h_{d}$ is a scalar. 
\begin{algorithm}[!t]
\caption{WGAN based training algorithm}\label{alg:wgan}
    \begin{algorithmic}[1]
        \REQUIRE the learning rate $\alpha$, the clipping parameter $c$, the batch size $m$, the number of iterations $n_{d}$, the domain classifier $h_{d}$ with parameters  $w_{d}$, the feature extractor $M_S$ with parameters $w_{M_{S}}$.\\
        \STATE \textbf{Initialization}: $w_{M_{T}} = w_{M_{S}}$
        \WHILE{$w_{M_{T}}$ not converged}
            \FOR{$i=1,\ldots,n_{d}$}
                \STATE Get $m$ samples $\{\mathbf{x}^{S}_{n}\}_{n=1}^{m}$ from $\mathbb{X}_{S}$
                \STATE Get $m$ samples $\{\mathbf{x}^{T}_{n}\}_{n=1}^{m}$ from $\mathbb{X}_{T}$
                \STATE $g_{w_{d}} \leftarrow \nabla_{w_{d}}[\frac{1}{m}\sum\limits_{n=1}^{m}h_{d}(M_{S}(\mathbf{x}^{S}_{n})) -$\\ $\quad\quad\quad\frac{1}{m}\sum\limits_{n=1}^{m}h_{d}(M_{T}(\mathbf{x}^{T}_{n}))]$\label{lst:alg:loss-critic}
                \STATE $w_{d} \leftarrow w_{d} - \alpha\;Opt(w_{d}, g_{w_{d}})$
                \STATE $w_{d} \leftarrow clip(w_{d}, c)$
            \ENDFOR
            \STATE Get $m$ samples $\{\mathbf{x}^{S}_{n}, \mathbf{y}_{n}\}_{n=1}^{m}$ from $(\mathbb{X}_{S}, \mathbb{Y}_{S})$
            \STATE Get $m$ samples $\{\mathbf{x}^{T}_{n}\}_{n=1}^{m}$ from $\mathbb{X}_{T}$
            \STATE $g_{w_{M_{T}}} \leftarrow \nabla_{w_{M_{T}}}[\frac{1}{m}\sum\limits_{n=1}^{m}h_{d}(M_{T}(\mathbf{x}_{n})) -$\\ $\quad\quad\quad\quad\frac{1}{m}\sum\limits_{n=1}^{m}\mathbf{y}_{n}^{T}\log(h^{\star}(M_{T}(\mathbf{x}_n)))]
                $\label{lst:alg:addition}
            \STATE $w_{M_{T}} \leftarrow w_{M_{T}} - \alpha\;Opt(w_{M_{T}}, g_{w_{M_{T}}})$
        \ENDWHILE\\
    \RETURN $M_{T}$\\
    \end{algorithmic}
\end{algorithm}
\vspace{-6pt}
\subsection{Source domain training}\vspace{-4pt}
As a first step, we optimize $M$ and $h$ using the labeled data $(\mathbf{x}_{s},\mathbf{y}_{s})$ from source domain $(\mathbb{X}_{S}, \mathbb{Y}_{S})$, where $\mathbf{y}_{S}$ is an 1-hot encoding of the available classes. We employ the binary cross-entropy as the source domain loss function $\epsilon_{S}(h, f_{S})$:
\begin{equation}\label{eq:label-error-method}
    \mathcal{L}_{\text{labels}}(h, M) = -\sum\limits_{(\mathbf{x},\mathbf{y})\in(\mathbb{X}_{S},\mathbb{Y}_{S})}\mathbf{y}^{T}\log(h(M(\mathbf{x})))\text{,}
\end{equation}
\noindent
and we obtain the classifier $h^{\star}$ and the source domain feature extractor $M_{S}$ by
\begin{equation}
    h^{\star}, M_{S} = \underset{h, M}{\text{argmin }}\mathcal{L}_{\text{labels}}(h, M)\text{.}
\end{equation}
The parameters of the feature extractor $M_S$ are denoted $w_{M_S}$ and will be used as initial values for the adapted feature extractor $M_T$.
%
%
\subsection{Wasserstein adversarial formulation}
As a second step, we aim at adapting $M_S$ to the target domain using an adversarial training process as described in Section~\ref{subsec:adversarialform}. However, as pointed out in Section~\ref{subsec:previous-work}, using $d_{\mathcal{H}\Delta\mathcal{H}}$ as a discrepancy measure between distributions yields computational problems. To alleviate this issue, we propose to employ the order-1 Wasserstein distance (called Wasserstein distance from now on) $W$~\cite{arjovsky:2017:icml,edwards:2011:math} as a metric for the discrepancy between $\mathcal{Z}_{S}$ and $\mathcal{Z}_{T}$. $W$ comes as a natural and intuitive candidate for our method since its minimization does not suffer from the problems mentioned in Section~\ref{subsec:previous-work}. Additionally, the usage of a binary label classifier $h$ by $\mathcal{H}\Delta\mathcal{H}$-distance introduces an intractable problem in practice, but $W$ distance does not suffer from this problem as it does not require $h$ to be a binary classifier~\cite{redko:2017:mlkdd}. Furthermore, $W$ is proven to enhance the results of adversarial training, and it is a weak topology over the space of $\mathcal{Z}$ allowing important convergence modes such as smooth convergence and point-wise convergence~\cite{ben-david:2010:mlrj,arjovsky:2017:icml,redko:2017:mlkdd}. Finally, $W$ accounts for the geometry of the space of $\mathcal{Z}$, therefore being an appropriate distance for measuring the discrepancy between the two distributions $\mathcal{Z}_{T}$ and $\mathcal{Z}_{T}$~\cite{redko:2017:mlkdd}. $W$ is defined~\cite{arjovsky:2017:icml} as 
\begin{equation}\label{eq:wasserstein-distance}
    W(\mathcal{Z}_{S}, \mathcal{Z}_{T})=\underset{(\mathbf{z}_{S}, \mathbf{z}_{T})\sim\prod{(\mathcal{Z}_{T}, \mathcal{Z}_{T})}}{\inf}\mathbb{E}_{(\mathbf{z}_{S}, \mathbf{z}_{T})}[||\mathbf{z}_{S}-\mathbf{z}_{T}||]\text{,}
\end{equation}
\noindent
where $\prod{(\mathcal{Z}_{S}, \mathcal{Z}_{T})}$ is the set of all joint distributions whose respective marginals are $\mathcal{Z}_{S}$ and $\mathcal{Z}_{T}$. It is proven~\cite{redko:2017:mlkdd} that we can use $W$ as a divergence metric to upper bound the target error $\epsilon_{T}$ using an expression similar to Eq.~\eqref{eq:upper_bound}:
\begin{equation}\label{eq:wasserstein-upper-bound}
    \epsilon_{T}(h, f_{T}) \leq \epsilon_{S}(h, f_{S})+W(\mathcal{Z}_{S}, \mathcal{Z}_{T})+\lambda\text{.}
\end{equation}
In Eq.~\eqref{eq:wasserstein-upper-bound} the factors corresponding to the source error and the ideal joint classifier (i.e., $\epsilon_{S}(h,f_{S})$ and $\lambda$) are the same as in Eq.~\eqref{eq:upper_bound}. The difference is the factor representing the discrepancy between the two distributions, $\mathcal{Z}_{S}$ and $\mathcal{Z}_{T}$, i.e., $W(\mathcal{Z}_{S}, \mathcal{Z}_{T})$ instead of $d_{H\Delta H}(\mathcal{Z}_{S}, \mathcal{Z}_{T})$. Thus, we can use the WGAN training algorithm presented in~\cite{arjovsky:2017:icml}, which yields the adapted feature extractor $M_{T}$. The process of the adaptation of $M_{T}$ is performed by the iterative minimization of the losses
\begin{align}
    &\sum\limits_{\mathbf{x}\in\mathbb{X}_{S}}h_{d}(M_{S}(\mathbf{x})) - \sum\limits_{\mathbf{x} \in \mathbb{X}_{T}}h_{d}(M_{T}(\mathbf{x}))\text{ and}\label{eq:wgan-loss-critic}\\
    &\sum\limits_{\mathbf{x} \in \mathbb{X}_{T} }h_{d}(M_{T}(\mathbf{x})) + \mathcal{L}_{\text{labels}}(h^{\star}, M_T) \text{. }\label{eq:extra-loss}
\end{align}
\noindent
Eq.~\eqref{eq:wgan-loss-critic} and~\eqref{eq:extra-loss} (without the $\mathcal{L}_{\text{labels}}(h^{\star}, M_T)$ term) are proposed in the original WGAN algorithm and their minimization is shown to minimize the $W$ distance in Eq.~\eqref{eq:wasserstein-distance}~\cite{arjovsky:2017:icml}. We enhance the original WGAN losses and training algorithm by including an extra cross-entropy loss (the second term in Eq.~\eqref{eq:extra-loss}), similarly to~\cite{gharib2018unsupervised}. The goal of this addition is to prevent the adapted model from deteriorating the performance on the source domain $\mathcal{D}_{S}$.
%
%
%
Our method is summarized in Algorithm~\ref{alg:wgan} and our enhancement to the original WGAN algorithm is at line~\ref{lst:alg:addition} of this algorithm. Lines~\ref{lst:alg:loss-critic} and~\ref{lst:alg:addition} are the calculations over a minibatch of the gradients of Eq.~\eqref{eq:wgan-loss-critic} and~\eqref{eq:extra-loss}, respectively. Note that Algorithm~\ref{alg:wgan} uses an optimizer $Opt$ chosen as the RMSProp optimizer~\cite{ruder:2016:arxiv} and a clipping function $clip$ defined as:
\begin{equation}
    clip(x, c) = \max ( \min ( x, c), -c).
\end{equation}
\begin{figure*}[!ht]
    \centering
    \subcaptionbox{Non adapted model\label{subfig:non-adapted}}
    {\includegraphics[width=.95\columnwidth]{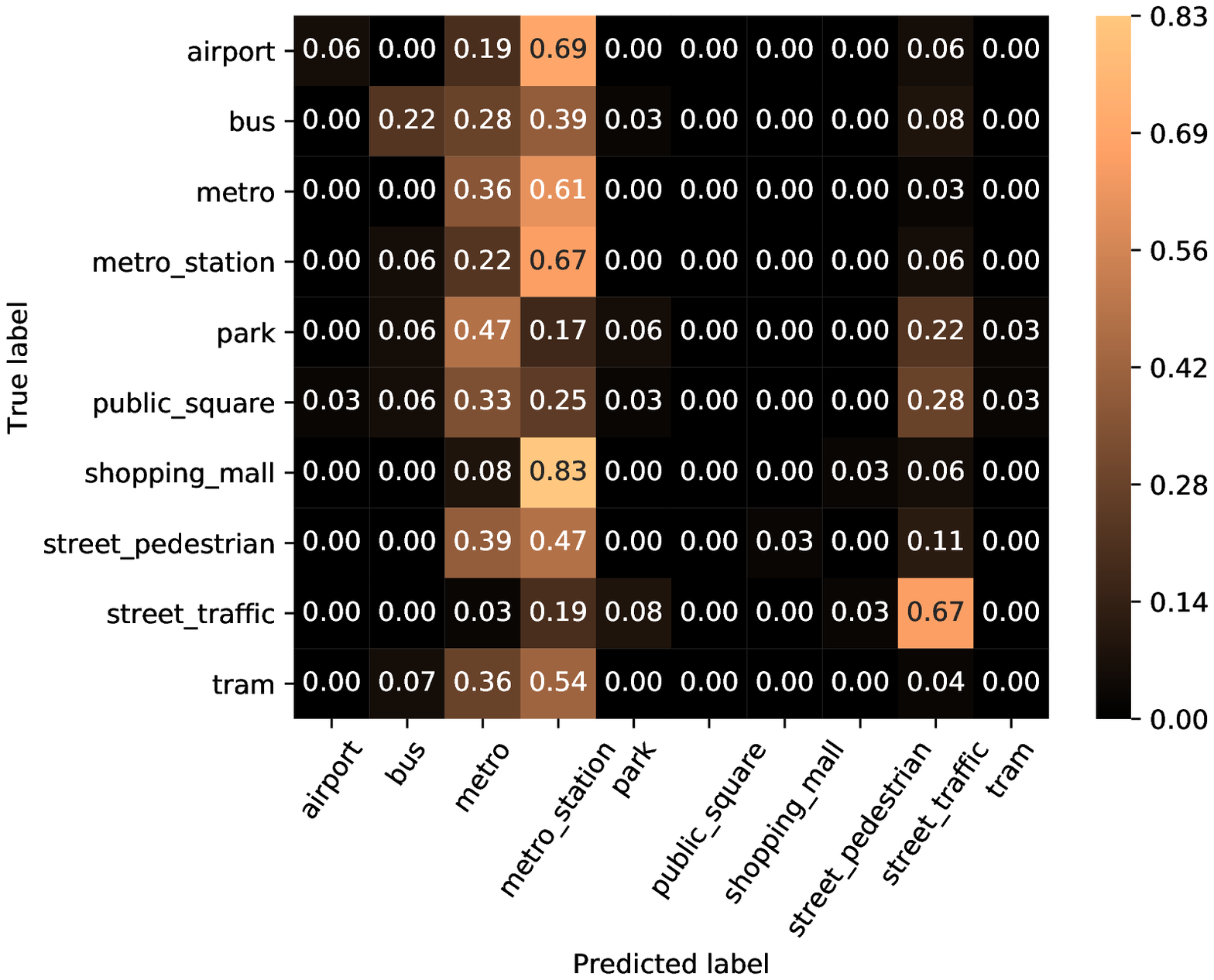}}\hspace{.5em}%
    \subcaptionbox{Adapted model\label{subfig:adapted}}
    {\includegraphics[width=.95\columnwidth]{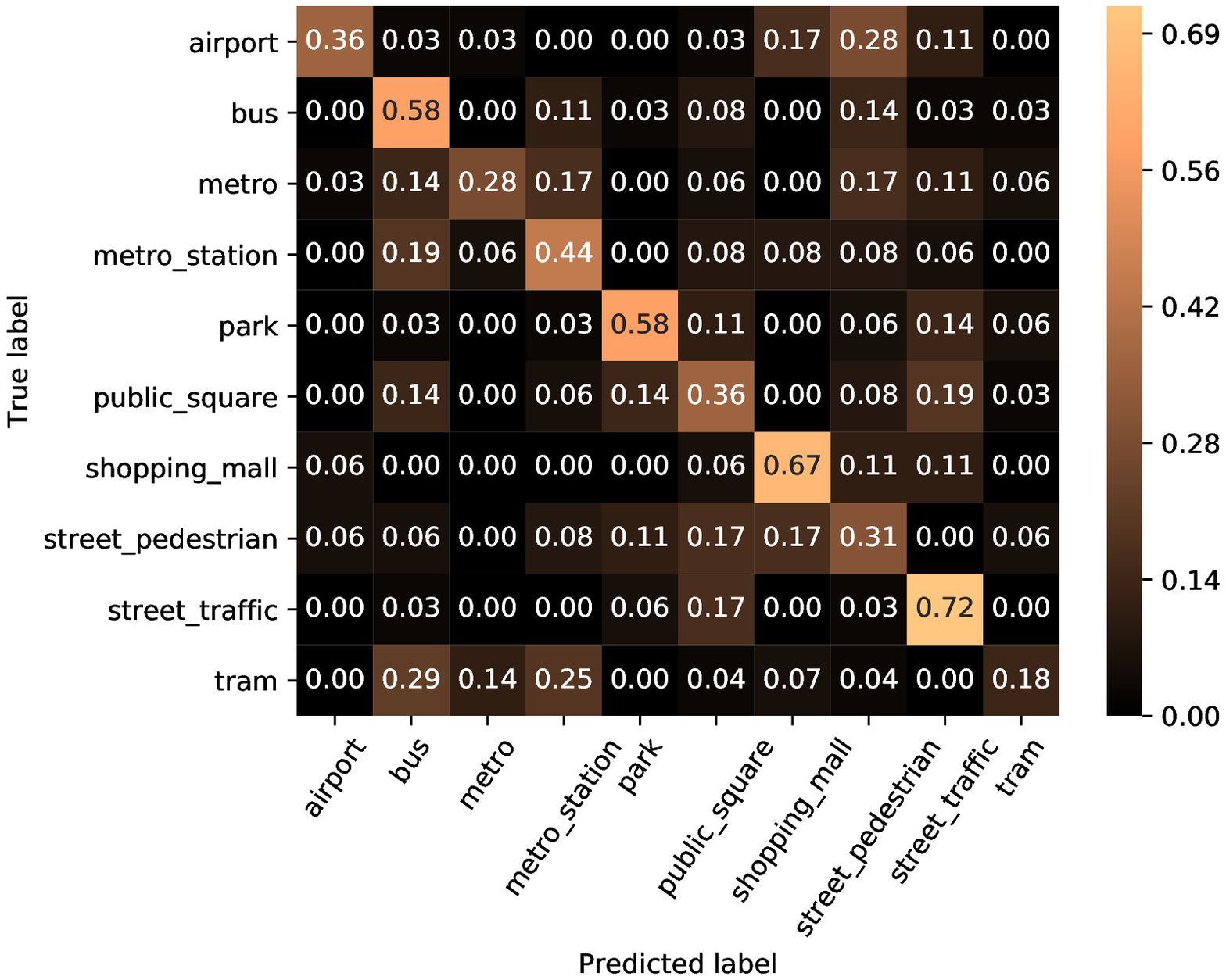}}
    \caption{Normalized confusion matrices for the non- and adapted models on the target domain data $\mathbf{x}_{T}\in\mathbb{X}_{T}$.}
    \label{fig:confusion-matrices}
\end{figure*}
%
%
\section{Evaluation setup}\label{sec:evaluation}\vspace{-6pt}
To evaluate our method and obtain comparable results, we use the same data and models for $M$ and $h$ as in the previous SOTA method for ASC unsupervised DA~\cite{gharib2018unsupervised}. The data that we use are from the TUT Acoustic Scenes dataset used in the previous SOTA approach~\cite{gharib2018unsupervised,gharib:2018zenodo-audasc}, and consist of 64 log mel-bank energies, extracted from audio recordings of 10 different acoustic scenes, namely: ``airport'', ``bus'', ``metro'', ``metro station'', ``park'', ``public square'', ``shopping mall'', ``street pedestrian'', ``street traffic'', and ``tram''. The recordings are performed with three different devices, denoted A, B, and C. Device A is regarded as a high quality recording device and the other two as consumer devices. Data from device A are 24 hours long, and data from devices B and C are two hours long, each. We follow the splitting of the data in training, validation, and testing as in~\cite{gharib2018unsupervised,gharib:2018zenodo-audasc}, leading to training and validation set consisting of approximately 15 hours of data from device A, 1.35 hours from device B, and 1.35 hours from device C. Consequently, the testing set consists of approximately 7 hours of data from device A, 0.5 hours from device B, and 0.5 hours from device C. We also adopt the definition of source and target domain data as in~\cite{gharib2018unsupervised,gharib:2018zenodo-audasc}. That is, we regard the data from device A as our source domain data, and the data from the consumer devices as our target domain data. 

We employ the non-adapted, pre-trained model referred to as ``Kaggle model'' in~\cite{gharib2018unsupervised} as our $M_{S}$, and the pre-trained classifier as our $h$. Both $M_{S}$ and $h$ are freely available at~\cite{gharib:2018zenodo-audasc}. $M_{S}$ consists of five 2D convolutional neural network (CNN) layers, with square kernel shapes of widths $\{11, 5, 3, 3, 3 \}$, and amount of channels $\{48, 128, 192, 192, 128\}$. The utilized stride is $(2, 3)$ for the first two CNNs and $(1, 1)$ for the others. All CNNs are followed by a rectified linear unit (ReLU) non-linearity, and the first two and last CNNs also use batch normalization and max pooling, with square kernels of shape $3$ and a stride of $\{(1, 2), (2, 2), (1, 2)\}$. $h$ consists of three feed-forward layers, each one followed by a ReLU. The output non-linearity of $h$ is the softmax function. The method was implemented using the PyTorch framework~\cite{paszke:2017:pytorch} for Python programming language. For the adaptation stage, the RMSProp optimizer is used with a learning rate of $5\times10^{-5}$ and have its rest parameters at the proposed default values. We use a batch size of $16$ and the adapted $M_{T}$ was obtained after the saturation of the first term in Eq.~\ref{eq:extra-loss} (after approximately 300 epochs). For a reproducible research purpose, we offer all of our code\footnote{\url{https://github.com/dr-costas/undaw}} and adapted $M_{T}$ model\footnote{\url{https://doi.org/10.5281/zenodo.2649151}} in online repositories.
%
%
\vspace{-10pt}
\section{Results and discussion}\label{sec:results}
\vspace{-7pt}
We present the mean accuracy and the normalized classification results per class as a confusion matrix, for adapted and non-adapted models. Table~\ref{tab:results} presents the obtained mean accuracy of the adapted and non-adapted models from our method and the previous SOTA approach. As it can be seen from Table~\ref{tab:results}, our adaptation method provides a significant increase of $13\%$ in mean accuracy for the target domain compared to the previous SOTA. The $1\%$ decrease in accuracy of the adapted model on the source domain and the $1\%$ increase in accuracy of the non-adapted model on the target domain, are considered insignificant and can be attributed to number rounding processes. In Figure~\ref{fig:confusion-matrices} are the confusion matrices for the non-adapted and adapted models on the target domain data $\mathbf{x}_{T}\in\mathbb{X}_{T}$.
\begin{table}
\centering
\caption{Mean accuracy of the adapted and non adapted models on the source $\mathcal{D}_{S}$ and target $\mathcal{D}_{T}$ domains.}
\label{tab:results}
\smallskip
\begin{tabular}{lll|ll}
 & \multicolumn{2}{c|}{\textbf{Non adapted}} & \multicolumn{2}{c}{\textbf{Adapted}} \\ \cline{2-5} 
  & $\mathcal{D}_{S}$ & $\mathcal{D}_{T}$ & $\mathcal{D}_{S}$ & $\mathcal{D}_{T}$ \\ 
Previous SOTA~\cite{gharib2018unsupervised} & \multicolumn{1}{c}{0.65} & \multicolumn{1}{c|}{0.20} & \multicolumn{1}{c}{0.65} & \multicolumn{1}{c}{0.32}\\
Proposed approach & \multicolumn{1}{c}{0.65} & \multicolumn{1}{c|}{0.21} & \multicolumn{1}{c}{0.64} & \multicolumn{1}{c}{$\mathbf{0.45}$}
\end{tabular}
\end{table}
In Figure~\ref{subfig:adapted} it can be seen that there is an increase of the values on the diagonal, compared to Figure~\ref{subfig:non-adapted}. This indicates that with the proposed approach, the discrepancy between $\mathcal{Z}_{S}$ and $\mathcal{Z}_{T}$ decreased and $h$ is able to perform classification up to an extent. There are classes where the adapted model yielded an accuracy above 50\%, like street traffic, shopping mall, bus, and park. For the same classes, the non-adapted model could only classify the street traffic class, while for the others it yielded an accuracy below 40\%. 
%
%
\vspace{-3pt}
\section{Conclusions and future work}\label{sec:conclusions}\vspace{-7pt}
In this work we presented a first approach for acoustic scene classification unsupervised domain adaptation that is based on the Wasserstein distance, along with the underlying theoretical framework. The presented method is evaluated on the TUT Acoustic Scenes dataset and the obtained results surpassed the previous state-of-the-art mean accuracy on the target domain by $13\%$. Future work on ASC DA will include the evaluation of the method on a larger dataset, along with using better ASC models. Besides, we will also apply these methods to alternative problems such as speech enhancement, where DA may help accounting for very adverse recording conditions.
%
%
\vspace{-6pt}
\section{Acknowledgement}\vspace{-7pt}
Part of the computations leading to these results were performed on a TITAN-X GPU donated by NVIDIA to K. Drossos. The authors wish to acknowledge CSC-IT Center for Science, Finland, for computational resources. The research leading to these results has received funding from the European Research Council under the European Union's H2020 Framework Programme through ERC Grant Agreement 637422 EVERYSOUND. P. Magron is supported by the Academy of Finland, project no. 290190.
\bibliographystyle{IEEEtran}
\bibliography{refs}

\begin{thebibliography}{10}
\providecommand{\url}[1]{#1}
\csname url@samestyle\endcsname
\providecommand{\newblock}{\relax}
\providecommand{\bibinfo}[2]{#2}
\providecommand{\BIBentrySTDinterwordspacing}{\spaceskip=0pt\relax}
\providecommand{\BIBentryALTinterwordstretchfactor}{4}
\providecommand{\BIBentryALTinterwordspacing}{\spaceskip=\fontdimen2\font plus
\BIBentryALTinterwordstretchfactor\fontdimen3\font minus
  \fontdimen4\font\relax}
\providecommand{\BIBforeignlanguage}[2]{{%
\expandafter\ifx\csname l@#1\endcsname\relax
\typeout{** WARNING: IEEEtran.bst: No hyphenation pattern has been}%
\typeout{** loaded for the language `#1'. Using the pattern for}%
\typeout{** the default language instead.}%
\else
\language=\csname l@#1\endcsname
\fi
#2}}
\providecommand{\BIBdecl}{\relax}
\BIBdecl

\bibitem{han:2017:dcase}
Y.~Han and J.~Park, ``Convolutional neural networks with binaural
  representations and background subtraction for acoustic scene
  classification,'' DCASE2017 Challenge, Tech. Rep., Sep 2017.

\bibitem{weiping:2017:dcase}
Z.~Weiping, Y.~Jiantao, X.~Xiaotao, L.~Xiangtao, and P.~Shaohu, ``Acoustic
  scene classification using deep convolutional neural network and multiple
  spectrograms fusion,'' DCASE2017 Challenge, Tech. Rep., Sep 2017.

\bibitem{sakashita:2018:dcase}
Y.~Sakashita and M.~Aono, ``Acoustic scene classification by ensemble of
  spectrograms based on adaptive temporal divisions,'' DCASE2018 Challenge,
  Tech. Rep., Sep 2018.

\bibitem{dorfer:2018:dcase}
M.~Dorfer, B.~Lehner, H.~Eghbal-zadeh, H.~Christop, P.~Fabian, and W.~Gerhard,
  ``Acoustic scene classification with fully convolutional neural networks and
  {I}-vectors,'' DCASE2018 Challenge, Tech. Rep., Sep 2018.

\bibitem{gharib2018unsupervised}
S.~Gharib, K.~Drossos, E.~\c{C}akir, D.~Serdyuk, and T.~Virtanen,
  ``Unsupervised adversarial domain adaptation for acoustic scene
  classification,'' in \emph{Proceedings of the Detection and Classification of
  Acoustic Scenes and Events 2018 Workshop (DCASE2018)}, Nov 2018, pp.
  138--142.

\bibitem{cortes:2008:alt}
C.~Cortes, M.~Mohri, M.~Riley, and A.~Rostamizadeh, ``Sample selection bias
  correction theory,'' in \emph{Algorithmic Learning Theory}, Y.~Freund,
  L.~Gy{\"o}rfi, G.~Tur{\'a}n, and T.~Zeugmann, Eds.\hskip 1em plus 0.5em minus
  0.4em\relax Berlin, Heidelberg: Springer Berlin Heidelberg, 2008, pp. 38--53.

\bibitem{ganin2016domain}
Y.~Ganin, E.~Ustinova, H.~Ajakan, P.~Germain, H.~Larochelle, F.~Laviolette,
  M.~Marchand, and V.~Lempitsky, ``Domain-adversarial training of neural
  networks,'' \emph{The Journal of Machine Learning Research}, vol.~17, no.~1,
  pp. 2096--2030, 2016.

\bibitem{torralba:2011:cvpr}
A.~{Torralba} and A.~A. {Efros}, ``Unbiased look at dataset bias,'' in
  \emph{{IEEE} Computer Vision and Pattern Recognition ({CVPR})}, Jun 2011, pp.
  1521--1528.

\bibitem{gretton:2009:chapter}
A.~Gretton, A.~Smola, J.~Huang, M.~Schmittfull, K.~Borgwardt, and
  B.~Sch{\"o}lkopf, \emph{Covariate shift and local learning by distribution
  matching}.\hskip 1em plus 0.5em minus 0.4em\relax Cambridge, {MA, USA}: MIT
  Press, 2009, pp. 131--160.

\bibitem{tzeng:cvpr:2017}
E.~Tzeng, J.~Hoffman, K.~Saenko, and T.~Darrell, ``Adversarial discriminative
  domain adaptation,'' in \emph{2017 {IEEE} Conference on Computer Vision and
  Pattern Recognition ({CVPR})}, Jul 2017, pp. 2962--2971.

\bibitem{Tommasi_GCPR_2015}
T.~Tommasi, N.~Patricia, B.~Caputo, and T.~Tuytelaars, ``A deeper look at
  dataset bias,'' in \emph{German Conference on Pattern Recognition}, ser.
  Lecture Notes in Computer Science, vol. 9358.\hskip 1em plus 0.5em minus
  0.4em\relax Springer International Publishing, Oct 2015, pp. 504--516.

\bibitem{ben-david:2010:mlrj}
S.~Ben-David, J.~Blitzer, K.~Crammer, A.~Kulesza, F.~Pereira, and J.~W.
  Vaughan, ``A theory of learning from different domains,'' \emph{Machine
  Learning}, vol.~79, no.~1, pp. 151--175, May 2010.

\bibitem{zhao2018multiple}
H.~Zhao, S.~Zhang, G.~Wu, J.~M.~F. Moura, J.~P. Costeira, and G.~J. Gordon,
  ``Adversarial multiple source domain adaptation,'' in \emph{Advances in
  Neural Information Processing Systems 31}, S.~Bengio, H.~Wallach,
  H.~Larochelle, K.~Grauman, N.~Cesa-Bianchi, and R.~Garnett, Eds.\hskip 1em
  plus 0.5em minus 0.4em\relax Curran Associates, Inc., 2018, pp. 8559--8570.

\bibitem{pei:2018:aaai}
Z.~Pei, Z.~Cao, M.~Long, and J.~Wang, ``Multi-adversarial domain adaptation,''
  in \emph{Proceedings of the 32nd AAAI Conference on Artificial Intelligence},
  2018.

\bibitem{hoffman:2018:jmlr}
J.~Hoffman, E.~Tzeng, T.~Park, J.-Y. Zhu, P.~Isola, K.~Saenko, A.~Efros, and
  T.~Darrell, ``{C}y{CADA}: Cycle-consistent adversarial domain adaptation,''
  in \emph{Proceedings of the 35th International Conference on Machine
  Learning}, ser. Proceedings of Machine Learning Research, J.~Dy and
  A.~Krause, Eds., vol.~80.\hskip 1em plus 0.5em minus 0.4em\relax PMLR, Jul
  2018, pp. 1989--1998.

\bibitem{chadha:2018:arxiv}
\BIBentryALTinterwordspacing
A.~Chadha and Y.~Andreopoulos, ``Improving adversarial discriminative domain
  adaptation,'' \emph{CoRR}, vol. abs/1809.03625, 2018. [Online]. Available:
  \url{http://arxiv.org/abs/1809.03625}
\BIBentrySTDinterwordspacing

\bibitem{redko:2017:mlkdd}
I.~Redko, A.~Habrard, and M.~Sebban, ``Theoretical analysis of domain
  adaptation with optimal transport,'' in \emph{Machine Learning and Knowledge
  Discovery in Databases}, M.~Ceci, J.~Hollm{\'e}n, L.~Todorovski, C.~Vens, and
  S.~D{\v{z}}eroski, Eds.\hskip 1em plus 0.5em minus 0.4em\relax Cham: Springer
  International Publishing, 2017, pp. 737--753.

\bibitem{arjovsky:2017:icml}
M.~Arjovsky, S.~Chintala, and L.~Bottou, ``{W}asserstein generative adversarial
  networks,'' in \emph{Proceedings of the 34th International Conference on
  Machine Learning}, ser. Proceedings of Machine Learning Research, D.~Precup
  and Y.~W. Teh, Eds., vol.~70.\hskip 1em plus 0.5em minus 0.4em\relax
  International Convention Centre, Sydney, Australia: PMLR, Aug 2017, pp.
  214--223.

\bibitem{gharib:2018zenodo-audasc}
\BIBentryALTinterwordspacing
S.~Gharib, K.~Drossos, E.~\c{C}akir, D.~Serdyuk, and T.~Virtanen, ``Adversarial
  unsupervised domain adaptation for acoustic scene classification,'' online:
  https://doi.org/10.5281/zenodo.1401995, Aug 2018. [Online]. Available:
  \url{https://doi.org/10.5281/zenodo.1401995}
\BIBentrySTDinterwordspacing

\bibitem{blitzer:2008:nips}
J.~Blitzer, K.~Crammer, A.~Kulesza, F.~Pereira, and J.~Wortman, ``Learning
  bounds for domain adaptation,'' in \emph{Advances in Neural Information
  Processing Systems 20}, J.~C. Platt, D.~Koller, Y.~Singer, and S.~T. Roweis,
  Eds.\hskip 1em plus 0.5em minus 0.4em\relax Curran Associates, Inc., 2008,
  pp. 129--136.

\bibitem{vapnik:1999:springer}
V.~N. Vapnik, \emph{The Nature of Statistical Learning Theory}, 2nd~ed., ser.
  Statistics for Engineering and Information Science.\hskip 1em plus 0.5em
  minus 0.4em\relax Springer, 1999.

\bibitem{ganin:2015:icml}
Y.~Ganin and V.~Lempitsky, ``Unsupervised domain adaptation by
  backpropagation,'' in \emph{Proceedings of the 32nd International Conference
  on Machine Learning (ICML)}, Nov 2015.

\bibitem{goodfellow:2014:nips}
I.~Goodfellow, J.~Pouget-Abadie, M.~Mirza, B.~Xu, D.~Warde-Farley, S.~Ozair,
  A.~Courville, and Y.~Bengio, ``Generative adversarial nets,'' in
  \emph{Advances in Neural Information Processing Systems 27}, Z.~Ghahramani,
  M.~Welling, C.~Cortes, N.~D. Lawrence, and K.~Q. Weinberger, Eds.\hskip 1em
  plus 0.5em minus 0.4em\relax Curran Associates, Inc., 2014, pp. 2672--2680.

\bibitem{edwards:2011:math}
D.~Edwards, ``On the kantorovich–rubinstein theorem,'' \emph{Expositiones
  Mathematicae}, vol.~29, no.~4, pp. 387--398, 2011.

\bibitem{ruder:2016:arxiv}
\BIBentryALTinterwordspacing
S.~Ruder, ``An overview of gradient descent optimization algorithms,''
  \emph{CoRR}, vol. abs/1609.04747, 2016. [Online]. Available:
  \url{http://arxiv.org/abs/1609.04747}
\BIBentrySTDinterwordspacing

\bibitem{paszke:2017:pytorch}
A.~Paszke, S.~Gross, and A.~Lerer, ``Automatic differentiation in {PyTorch},''
  in \emph{NIPS-W}, 2017.

\end{thebibliography}
\end{document}